\documentclass[aps,pra,twocolumn,superscriptaddress,showpacs]{revtex4-1}

\usepackage{graphicx}
\usepackage{amsmath}
\usepackage{comment}
\usepackage[colorlinks,linkcolor=blue,citecolor=blue]{hyperref}

\newcommand{\eqn}[1]{
\begin{eqnarray}
	#1
\end{eqnarray}
}

\begin{document}


\title{Quantum critical behavior influenced by measurement backaction in ultracold gases}


\author{Yuto Ashida}
\affiliation{Department of Physics, University of Tokyo, 7-3-1 Hongo, Bunkyo-ku, Tokyo
113-0033, Japan}
\author{Shunsuke Furukawa}
\affiliation{Department of Physics, University of Tokyo, 7-3-1 Hongo, Bunkyo-ku, Tokyo
113-0033, Japan}
\author{Masahito Ueda}
\affiliation{Department of Physics, University of Tokyo, 7-3-1 Hongo, Bunkyo-ku, Tokyo
113-0033, Japan}
\affiliation{RIKEN Center for Emergent Matter Science (CEMS), Wako, Saitama 351-0198, Japan
}

\date{\today}

\begin{abstract} 
Recent realizations of quantum gas microscope offer the possibility of continuous monitoring of the dynamics of a quantum many-body system at the single-particle level.  
By analyzing effective non-Hermitian Hamiltonians of interacting bosons in an optical lattice and continuum, we demonstrate that the backaction of quantum measurement shifts the quantum critical point and gives rise to a unique critical phase beyond the terrain of the standard universality class. We perform mean-field and strong-coupling-expansion analyses and show that  non-Hermitian contributions shift the superfluid--to-Mott-insulator transition point. Using a low-energy effective field theory, we discuss critical behavior of the one-dimensional interacting Bose gas subject to the measurement backaction. We derive an exact ground state of the effective non-Hermitian   Hamiltonian and find a unique critical behavior beyond the Tomonaga-Luttinger liquid universality class.  We propose experimental implementations of post-selections using quantum gas microscopes to simulate the non-Hermitian dynamics and argue that  our results can be investigated with current experimental techniques in ultracold atoms.
\end{abstract}


\maketitle
\section{Introduction}
Quantum gas microscopy has revolutionalized our approach to quantum many-body physics. A large number of atoms trapped in an optical lattice can now be probed at the single-atom level with the diffraction-limited spatial resolution \cite{BWS09,SJF10,MM15,CLW15,PMF15,EH15,OA15,EGJ15,RY16}. One can perform single-shot measurements of quantum many-body systems at an unprecedented precision for studies of strongly correlated systems  \cite{EM11,EM13,FT15,IR15,YA16,YA162}.
While many-body dynamical phenomena subject to measurement backaction have  been observed in ultracold atom experiments by using, for example, a low-resolution imaging \cite{PYS15} or a cavity \cite{BF13}, recent developments \cite{PPM152,AA15} in quantum gas microscopy have opened up the possibility of continuous monitoring of the many-body dynamics at the single-particle level \cite{YA15}.
Meanwhile, at such ultimate resolution, the measurement backaction is expected to be significant. One would naturally be led to the question of how the quantum critical behavior is modified due to the backaction of high-precision measurement.

Recently, unique dynamical aspects of many-body systems under the measurement backaction have been studied in various situations \cite{MKP14,LMD14,DS15,DS16,WAC15,WAC16,MG16}. The measurement backaction can be described as a sudden jump of a quantum state when the signal is detected. In contrast, if the system is continuously monitored and conditioned on a null-measurement outcome, it obeys the dynamics described by a non-Hermitian Hamiltonian \cite{DJ92,HC93,DAJ14}. For such continuously monitored systems, an exotic steady-state transition has been found in the spin chain model subject to spontaneous decay \cite{LTE14}, and, the atom-cavity system in the quantum Zeno regime has been analyzed \cite{KW15}.
However, it remains to be understood how the conventional notions of quantum phase transitions and universality in quantum critical phenomena  can be extended in the presence of continuous monitoring. Statistical mechanics would provide an answer, were it not for the measurement backaction. The frontier of quantum gas microscopy thus motivates us to elucidate its influence on the quantum criticality beyond the standard framework of statistical mechanics.


\begin{figure}[b]
\includegraphics[width=86mm]{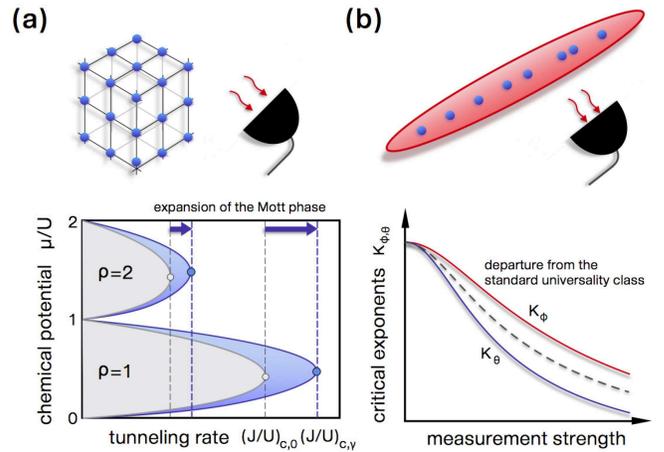} 
\caption{\label{fig1}
(color online). Schematic illustration of continuous observation of quantum critical phenomena in ultracold atoms loaded in (a) an optical lattice or (b) a one-dimensional trap. The measurement backaction (a) expands the Mott lobe and shifts the quantum critical point, and (b) gives rise to a unique critical behavior described by two critical exponents $K_{\phi}$, $K_{\theta}$ (solid curves) in sharp contrast with the standard universality class described by the single Tomonaga-Luttinger liquid parameter (dashed curve). Here $\mu$, $J$, $U$ and $\rho$ are the chemical potential, the hopping amplitude, the strength of the on-site interaction and the filling fraction, respectively.
}
\end{figure}

In this paper, we investigate unique roles played by the measurement backaction in quantum critical systems subject to continuous observation. We employ an effective non-Hermitian description and show that the influence of measurement backaction on eigenenergies and eigenstates of a system lead to significant changes in the quantum critical points and critical exponents (Fig.\;\ref{fig1}). In particular, we derive the formula that characterizes the measurement-induced shift of a quantum critical point of the superfluid--to-Mott-insulator transition. We also investigate the influence of the measurement backaction on the quantum critical phase in a one-dimensional system, and analytically find new critical exponents that depend on the strength of the measurement, indicating a unique critical behavior beyond the terrain of the standard universality class.  The formulation can straightforwardly be generalized to other physical systems subject to various types of external observations.

This paper is organized as follows. In Sec. \ref{modelsec}, we introduce  an effective non-Hermitian Hamiltonian as a model to describe a quantum many-body system subject to measurement backaction of continuous monitoring. In Sec. \ref{shiftsec}, we consider ultracold atoms trapped in an optical lattice and show that measurement backaction shifts the superfluid--to-Mott-insulator transition point. In Sec. \ref{criticalsec}, we consider critical behavior of a one-dimensional interacting Bose gas subject to measurement backaction and find that a unique critical behavior beyond the Tomonaga-Luttinger liquid class. In Sec. \ref{expsec}, we discuss experimental implementations of our models. In Sec. \ref{consec}, we conclude this paper.

\section{Effective non-Hermitian Hamiltonians \label{modelsec}}
We consider a quantum many-body system whose Hamiltonian $\hat{H}$ exhibits quantum critical behavior, and assume that the system is subject to continuous backaction of a general measurement characterized by a set of measurement operators $\{\hat{M}_{i}\}$. Since our primary aim is to elucidate the influence of the measurement backaction due to continuous observation, we consider the following effective non-Hermitian Hamiltonian:
\eqn{\label{nonher}
\hat{H}_{\rm eff}=\hat{H}-\frac{i\gamma}{2}\sum_{i}\hat{M}^{\dagger}_{i}\hat{M}_{i},
}
where the last term describes the measurement backaction with $\gamma$ characterizing the strength of the measurement. 
The non-Hermitian Hamiltonian (\ref{nonher}) can be obtained under a situation in which the system is continuously monitored and a null measurement outcome is post-selected \cite{DJ92,HC93,DAJ14}. We will discuss experimental conditions to realize such post-selection by using  quantum gas microscopy in Sec. \ref{expsec}. The non-Hermitian description has proved instrumental for a wide variety of open quantum systems \cite{IR09,CM69,GB69,CHL91,SJ95,HJS07,DC01,CY10,SR05,CH07,BC98,RI05,GT15,ZB15,LTE14,KW15}. 
The effective Hamiltonian in Eq. (\ref{nonher}), in general, has complex eigenvalues whose real  part describes the  energy and imaginary part gives the rate at which the corresponding eigenstate decays out of the Hilbert space of the system. 

Let us first consider a situation in which the many-body Hamiltonian commutes with all the measurement operators, i.e.,  $[\hat{H},\hat{M}_{i}]=0$ for $^{\forall} i$. In this case,  the second term in Eq. (\ref{nonher}) makes a contribution to the imaginary part of the eigenspectrum, leading to the decay of the state, however, the real parts of the eigenenvalues and the corresponding eigenstates remain unchanged. Therefore, there are no qualitative changes in physical properties of the critical behavior. In contrast, if the original Hamiltonian does not commute with some of the measurement operators, i.e., $\exists i$ such that $[\hat{H},\hat{M}_{i}]\neq0$, the  measurement backaction can influence (i) the real part of the eigenvalues and (ii) the eigenstates of Eq. (\ref{nonher}), which respectively lead to (i) the shift of the quantum critical point and (ii) a change in the critical exponent as shown below.

In the present paper, we focus our attention on the properties of an effective ground state, which is defined as the eigenstate corresponding to the lowest real part of the eigenspectrum. Such a state is found to be relevant in the non-Hermitian dynamics since it also has the minimal decay rate and thus survives longest in the time evolution in the systems considered in this paper. We also note that our model is different from a dissipative model described by a master equation, where one expects that the dissipative process eventually destroys subtle correlations underlying the quantum critical behavior, thereby leading to a high-temperature mixed steady state. Indeed, recent works have suggested that such steady states exhibit static properties similar to classical thermal equilibrium systems \cite{MA06,SLM13,TUC14} and infinite-temperature states \cite{YY14,SJ14}. In contrast, we show that continuous observation can sustain the quantum critical behavior and gives rise to unique phenomena due to the measurement backaction. 

\section{Measurement-induced shift of the quantum critical point\label{shiftsec}}
\subsection{Mean-field analysis}
We first consider ultracold atoms in an optical lattice. The system is described by the Bose-Hubbard Hamiltonian \cite{FM89}: 
\eqn{
\hat{H}_{\rm BH}&=&\hat{H_{0}}+\hat{V},\\
\hat{H}_{0}&=&\frac{U}{2}\sum_{i}\hat{n}_i(\hat{n}_i-1)-\mu\sum_{i}\hat{n}_{i},\\
\hat{V}&=&-J\sum_{\langle i,j\rangle}(\hat{b}^{\dagger}_{i}\hat{b}_{j}+{\rm H.c.}).
}Here $J$ and $U$ are the hopping amplitude and the strength of the on-site interaction, respectively,  $\mu$ is the chemical potential, $\hat{b}_{i}^{\dagger}$ and $\hat{b}_{i}$ are the creation and annihiliation operators of a boson at site $i$, and $\hat{n}_{i}\equiv\hat{b}^{\dagger}_{i}\hat{b}_{i}$. If  $J\ll U$, the ground state of the system remains to be in the gapped Mott insulator phase. With increasing $J/U$, the  energy gap decreases and  closes at a quantum critical point $(J/U)_{c}$, where the quantum phase transition to the superfluid phase occurs \cite{SS01}. The critical value $(J/U)_{c}$ corresponds to the tip of the Mott lobe with an integer filling $\rho$ in the $\mu-J$ phase diagram as schematically illustrated in Fig.\:\ref{fig1}a. For simplicity, we focus on the case of $\rho=1$ below. 

Let us consider a general measurement process by introducing an effective non-Hermitian Hamiltonian $\hat{H}_{\rm eff}$ in Eq. (\ref{nonher}).  An implementation of post-selection to simulate the non-Hermitian dynamics depends on the underlying dynamical process. For example, for a system subject to inelastic two-body loss of atoms ($\hat{M}_{i}=\hat{b}_{i}^{2}$), the post-selection of null quantum jump can be realized by using quantum gas microscopy and selecting those realizations in which the total number of particles in the initial state and agrees with that in the final state (see Sec. \ref{expsec} for details). The fidelity of such a process can be very high in view of recent developments \cite{BWS09,SJF10,MM15,CLW15,PMF15,EH15,OA15,EGJ15,RY16,EM11,EM13,FT15,PPM152,AA15,YA15} in achieving the near-unit fidelity of quantum gas microscopy as detailed in Sec. \ref{expsec}.

Unless all the operators $\hat{M}_{i}$ commute with $\hat{H}_{\rm BH}$, the measurement backaction shifts the real parts of eigenenergies. Such a measurement-induced shift can manifest itself as a shift in the quantum critical point. To show this, let us first perform a mean-field analysis \cite{FM89}. Putting $\hat{b}_{i}=\beta+\delta\hat{b}_{i}$ in the hopping term $\hat{V}$ and neglecting the second-order terms in $\delta\hat{b}_{i}$, we obtain the decoupled effective Hamiltonian 
$\hat{H}^{\rm MF}_{\rm eff}=\hat{V}^{\rm MF}+\hat{H}_{0}-(i\gamma/2)\sum_{i}\hat{M}_{i}^{\dagger}\hat{M}_{i}$, where $\hat{V}^{\rm MF}\equiv-Jz\sum_{i}(\beta^{*}\hat{b}_{i}+\beta\hat{b}_{i}^{\dagger}-|\beta|^2)$. The effective ground-state energy $E_{\beta,\gamma}$ is given by the real part of the eigenvalue of the ground state of $\hat{H}_{\rm eff}^{\rm MF}$, and it can be expanded as $E_{\beta,\gamma}=a_{0}+a_{2}(\gamma)|\beta|^2+a_{4}(\gamma)|\beta|^4+\cdots$. The coefficient $a_{2}(\gamma)$ is determined from the second-order perturbation in $\hat{V}^{\rm MF}$ for the Mott insulator state, and the phase boundary $(J/U)_{\gamma}$ is determined from the condition $a_{2}(\gamma)=0$.
 The critical point $(J/U)_{c,\gamma}$ corresponding to the tip of the Mott lobe can be determined from $\partial (J/U)_{\gamma}/\partial\mu =0$ under the condition $\partial^2 (J/U)_{\gamma}/\partial\mu^2 <0$. The relative amount of the measurement-induced shift of the quantum critical point $\Delta_{c,\gamma}\equiv ((J/U)_{c,\gamma}-(J/U)_{c,0})/(J/U)_{c,0}$ is then found to be

\eqn{\label{shift}
\Delta_{c,\gamma}&=&\frac{2+\sqrt{2}}{2}\left(\sqrt{2}c_{M,0}^{2}+c_{M,1}^{2}\right)\left(\frac{\gamma}{U}\right)^{2}\nonumber\\
\!&+&\!\frac{5}{16}(4\!+\!3\sqrt{2})\left(2c_{M,0}^{2}\!-\!c_{M,1}^{2}\right)^{2}\left(\frac{\gamma}{U}\right)^{4}
\!+O\left(\left(\frac{\gamma}{U}\right)^{6}\right)\!.\nonumber\\
}
Here we introduce the coefficients $c_{M,\rho}\equiv (\langle \rho+1|\hat{M}^{\dagger}\hat{M}|\rho+1\rangle-\langle \rho|\hat{M}^{\dagger}\hat{M}|\rho\rangle)/2$, where the site index $i$ is omitted (for example, $c_{M,\rho}=\rho$ for $\hat{M}=\hat{b}^2$). Equation (\ref{shift}) shows that, for a general measurement, the quantum critical point shifts in favor of the Mott phase due to the measurement backaction. Physically, this is due to the suppression of the hopping by the continuous quantum Zeno effect \cite{BA00,FP02,DAJ09,ZB14,YB13,SN08}, although we here focus on the influence of a relatively small measurement strength. The quantum phase transition discussed here is different from the one in Ref. \cite{LTE14}, where the steady-state transition occurs at which the gap in the imaginary part of the eigenenergies closes. We note that the ground state discussed here survives longest in the non-Hermitian dynamics, as demonstrated by the strong-coupling expansion analysis below. Thus, our mean-field analysis can be applied if a duration of the time evolution is long enough so that the initial state relaxes to the effective ground state.

\begin{figure}[t]
\includegraphics[scale=0.15]{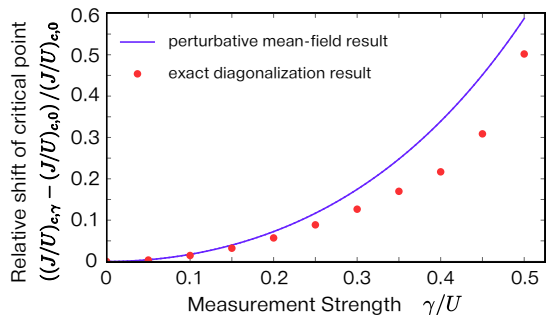}
\caption{\label{fig2}
(color online). Measurement-induced  shift $\Delta_{c,\gamma}$ of the quantum critical point plotted against the strength $\gamma/U$ of the parity measurement ($c_{M,\rho}=\rho$). The blue solid curve shows the perturbative mean-field result  in Eq. (\ref{shift}), and dots show the numerical results obtained by the exact diagonalization of the effective Hamiltonian (see the text). 
}
\end{figure} 

While the mean-field analysis may not give an accurate value of the transition point $(J/U)_{c,\gamma}$  \cite{DSF09}, we expect that the normalized relative shift $\Delta_{c,\gamma}$ should capture the right tendency of the measurement-induced shift. In particular, the perturbative formula (\ref{shift}) should be valid for small $\gamma/U$, as numerically supported as shown in Fig.\;\ref{fig2}. We here perform the exact diagonalization of the non-Hermitian Hamiltonian with the infinite-range hopping of amplitude $J/N$ among $N$ sites. The critical point is determined by first identifying the point at which the energy gap above the effective ground state becomes minimal and then by extrapolating the data to the thermodynamic limit. 
The numerical results asymptotically agree with Eq. (\ref{shift}) for small $\gamma/U$.

\subsection{Strong-coupling-expansion analysis}

Near integer values of $\mu/U$ in the $\mu-J$ diagram (Fig.\;\ref{fig1}a), fluctuations in the atom density are enhanced, so that  we need to consider a statistical mixture of Mott states with different fillings. To investigate these regimes, we perform the strong-coupling-expansion analysis \cite{FJK96,MF11} that gives asymptotically exact results for small $J/U$. 

\begin{figure}[t]
\includegraphics[width=86mm]{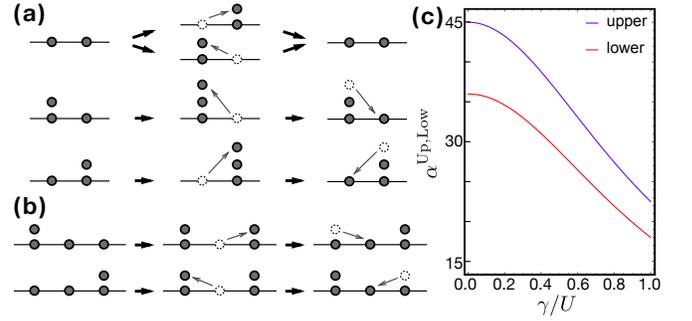}
\caption{\label{fig3}
Virtual processes relevant to the second-order strong-coupling-expansion analysis of a one-dimensional $\rho=1$ Mott phase, contributing to the (a) diagonal and (b) off-diagonal matrix elements. The filled circles indicate particles that occupy the lattice sites. The dashed circles indicate holes, from where particles move out due to hopping processes.  (c) Coefficients $\alpha_{d,\rho,\gamma}^{\rm Up,Low}$ for the upper (blue) and lower (red) boundaries of the Mott lobe  plotted against the measurement strength $\gamma/U$ for the case of the parity measurement ($c_{M,\rho}=\rho$) with $d=3$ and $\rho=1$.
}
\end{figure}

Let us consider a $d$-dimensional hypercubic lattice and analyze the Mott lobes with integer filling $\rho=1,2,\ldots$. For the unperturbed effective Hamiltonian
\eqn{
\hat{H}_{0}&=&\frac{U}{2}\sum_{i}\hat{n}_{i}(\hat{n}_{i}-1)-\mu\sum_{i}\hat{n}_{i}-\frac{i\gamma}{2}\sum_{i}\hat{M}^{\dagger}_{i}\hat{M}_{i},
}
the ground state with the lowest real energy remains to be the Mott state with  $\rho$ bosons on each site for $(\rho-1)U<\mu<\rho U$.  The first excited states near the upper phase boundary ($\mu\simeq\rho U$) constitute a family of degenerate states in which only a single site is occupied by $(\rho+1)$ particles and all the other sites are occupied by $\rho$ particles. In contrast, near the lower phase boundary ($\mu\simeq(\rho-1)U$), the first excited states consist of degenerate states in which only a single site is occupied by $(\rho-1)$ particles and all the other sites are occupied by $\rho$ particles. We apply a degenerate perturbation theory  \cite{MF11} to the ground and first excited states up to second order in the hopping term $\hat{V}=-J\sum_{\langle i,j\rangle}\left(\hat{a}_{i}^{\dagger}\hat{a}_{j}+{\rm H.c.}\right)$, and calculate their complex eigenvalues. Here, we take into account all possible processes in which a state in the low-energy manifold is virtually excited to a high-energy state, and then returns back to the manifold by the operations of $\hat{V}$; see Figs. \ref{fig3}(a) and (b). The real parts of the obtained eigenvalues are interpreted as the effective energies of the states. Then, the energy gaps, which are defined as the differences in real energies between the ground state and the first excited states, are calculated to be
\eqn{
\Delta_{d,\rho}^{\rm Up} &\!=\!& \!-\!2d(\rho+1)J\!+\!\rho U\!-\!\mu
\!\nonumber\\
&-&\frac{2d\rho(\rho+1)(2d-3)J^2}{U+(\gamma^2/U)(c_{M,\rho-1}-c_{M,\rho})^2}\nonumber\\
&-&\frac{d\rho(\rho+2)J^2}{U+(\gamma^2/4U)(c_{M,\rho-1}-c_{M,\rho+1})^2},
}
\eqn{
\Delta_{d,\rho}^{\rm Low} &\!=\!& \!-\!2d\rho J\!-\!(\rho-1)U\!+\!\mu\!\nonumber\\
&-&\!\frac{2d\rho(\rho+1)(2d-3)J^2}{U+(\gamma^2/U)(c_{M,\rho-1}-c_{M,\rho})^2}\nonumber\\
&-&\frac{d(\rho-1)(\rho+1)J^2}{U+(\gamma^2/4U)(c_{M,\rho-1}-c_{M,\rho+1})^2},
}
where $\Delta^{\rm Up}$ and $\Delta^{\rm Low}$ denote the energy gaps near the upper and lower boundaries of the Mott lobe. The phase boundary of the Mott phase can be identified as the point at which the energy gap closes:
\eqn{\label{cu}
\left(\frac{\mu}{U}\right)^{\rm Up}_{c,\gamma}-\rho=-2d(\rho+1)\frac{J}{U}-\alpha_{d,\rho,\gamma}^{\rm Up}\left(\frac{J}{U}\right)^2,
}
\eqn{\label{cl}
\left(\frac{\mu}{U}\right)^{\rm Low}_{c,\gamma}-(\rho-1)=2d\rho\frac{J}{U}+\alpha_{d,\rho,\gamma}^{\rm Low}\left(\frac{J}{U}\right)^2,
}
where the coefficients $\alpha_{d,\rho,\gamma}^{\rm Up,Low}$ are given by
\eqn{
\alpha_{d,\rho,\gamma}^{\rm Up}&=&\frac{2d\rho(\rho+1)(2d-3)}{1+(\gamma/U)^2(c_{M,\rho-1}-c_{M,\rho})^2}\nonumber\\
&+&\frac{d\rho(\rho+2)}{1+(\gamma/2U)^2(c_{M,\rho-1}-c_{M,\rho+1})^2},
}
\eqn{
\alpha_{d,\rho,\gamma}^{\rm Low}&=&\frac{2d\rho(\rho+1)(2d-3)}{1+(\gamma/U)^2(c_{M,\rho-1}-c_{M,\rho})^2}\nonumber\\
&+&\frac{d(\rho-1)(\rho+1)}{1+(\gamma/2U)^2(c_{M,\rho-1}-c_{M,\rho+1})^2}.
}
The presence of the backaction $\gamma>0$ decreases these coefficients $\alpha$ (see Fig. \ref{fig3}(c)), allowing an effective expansion of the Mott lobe as indicated in Eqs. (\ref{cu}) and (\ref{cl}), and as schematically illustrated in Fig. \ref{fig1}(a). We note that our results reproduce the known results \cite{FJK96} in the limit of $\gamma\to 0$.

Finally, we note that the ground state, i.e., the eigenstate that has the lowest real part of the eigenvalue, has the minimal decay rate (the longest lifetime) in the post-selected dynamics. For the sake of concreteness, let us assume $\rho=1$, $d=3$ and let $\Gamma_{e}$ and $\Gamma_{g}$ be the decay rate of the first excited  state and the ground state. We note that the decay rate is equal to the modulus of the imaginary part of the eigenvalue in our notation. Then, the strong-coupling-expansion analysis gives
\eqn{
\Gamma_{e}^{\rm Up}-\Gamma_{g}^{\rm Up} &=&\frac{36\gamma J^2}{U^2+\gamma^2(c_{M,\rho-1}-c_{M,\rho})^2}\nonumber\\
&+&\!\!\frac{9\gamma J^2}{U^2\!\!+\!\!(\gamma^2/4)(c_{M,\rho-1}-c_{M,\rho+1})^2}\!\!>\!\!0,\\
\Gamma_{e}^{\rm Low}-\Gamma_{g}^{\rm Low} &=&\frac{36\gamma J^2}{U^2+\gamma^2(c_{M,\rho-1}-c_{M,\rho})^2}>0,
}
which show that the decay occurs faster for the excited state than the ground state. Here $\Gamma^{\rm Up}$ and $\Gamma^{\rm Low}$ denote the decay rate calculated near the upper and lower boundaries of the Mott lobe, respectively. We can straightforwardly generalize the calculations to the higher excited states and show that the ground state considered here indeed has the minimal decay rate or the longest lifetime, among all the eigenstates of the effective non-Hermitian Hamiltonian.

\section{Influence on the critical behavior\label{criticalsec}}
\subsection{Model}
Let us now discuss a physical consequence of the measurement backaction on eigenstates in a quantum critical regime. To be specific, we consider a one-dimensional (1D) interacting Bose gas (Fig.\;\ref{fig1}b) described by the Lieb-Liniger Hamiltonian: 
\eqn{\hat{H}&=&\int dx\Biggl[\frac{-\hbar^2}{2m}\hat{\Psi}^{\dagger}(x)\frac{\partial^2}{\partial x^2}\hat{\Psi}(x)\nonumber\\
&&\;\;\;\;\;\;\;\;\;\;\;\;\;\;\;\;\;\;\;+\frac{g}{2}\hat{\Psi}^{\dagger}(x)\hat{\Psi}^{\dagger}(x)\hat{\Psi}(x)\hat{\Psi}(x)\Biggr],\label{HLL}} 
where $\hat{\Psi}(x)$ is the bosonic field operator and $m$ is the atomic mass. The repulsive interaction strength $g>0$ is given by $g=2\hbar\omega a_{r}$, where $\omega$ is the transverse confining frequency, and $a_{r}$ is the elastic scattering length. The low-energy critical behavior of this system is effectively described by the Tomonaga-Luttinger liquid (TLL) Hamiltonian \cite{HFD81}: $\hat{H}_{\rm TLL}=\int dx\hbar/(2\pi)[v_{J}(\partial_{x}\hat{  \theta})^2+v_{N}(\partial_{x}\hat{ \phi})^2]$, where the bosonic fields $\hat{  \theta}$ and $\hat{ \phi}$ satisfy $[\partial_{x}\hat{ \phi}(x),\hat{  \theta}(x')]=i\pi\delta(x-x')$, $v_{J}$ is the phase stiffness, and $v_{N}$ is the density stiffness. Here $\hat{  \theta}$ is related to the phase of the bosonic field operator as 
$\hat{\Psi}^{\dagger}(x)=\sqrt{\hat{\rho}(x)}e^{-i\hat{  \theta}(x)}$,
and $\hat{ \phi}$ is related to the density operator $\hat{\rho}(x)$ as $\hat{\rho}(x)\simeq [\rho_{0}+\partial_{x}\hat{ \phi}(x)/\pi]\sum_{p=-\infty}^{\infty} e^{2ip(\pi\rho_{0}x+\hat{ \phi}(x))}$, where $\rho_{0}$ is the average density. In the TLL, various correlation functions decay algebraically with exponents determined by the single TLL parameter  $K=\sqrt{v_{J}/v_{N}}$. In the Lieb-Liniger model, the Galilean invariance ensures the relation $v_{J}=\hbar\pi\rho_{0}/m$ \cite{CMA11}, and $v_{N}$ taks the following asymptotic forms \cite{BHP03,MAC04}:
\eqn{\label{asp}
v_{N}=\begin{cases}
\frac{v_{J}u}{\pi^2}\left(1-\frac{\sqrt{u}}{2\pi}\right) & {\rm for}\;\; u\ll 1;\\
v_{J}\left(1-\frac{8}{u}+O\left(\frac{1}{u^2}\right)\right)& {\rm for}\;\;u\gg 1,
\end{cases}
}
where $u\equiv mg/(\hbar^2\rho_{0})$ is the normalized strength of the interaction.

Let us discuss how the measurement backaction alters the  Hamiltonian of the system. For the sake of concreteness, we consider a system subject to  inelastic two-body loss of atoms. In such a situation, the backaction gives a non-Hermitian contribution to the interaction term, leading to the replacement $g\to g-i\gamma$ in Eq. (\ref{HLL}), where $\gamma=2\hbar\omega a_{i}$ characterizes the measurement strength determined from the inelastic scattering length $a_{i}$ \cite{SN08,JJGR09}.  Accordingly, as inferred from the analytic continuation of Eq. (\ref{asp}), the measurement backaction renormalizes the density stiffness $v_{N}$ to a complex value $\tilde{v}_{N}(\gamma)e^{-i\delta_{\gamma}}$, where $\tilde{v}_{N}$ and $\delta_{\gamma}$ are the real parameters that depend on the measurement strength $\gamma$. We thus arrive at the following effective non-Hermitian Hamiltonian: 

\eqn{\label{effective}
\hat{H}_{\rm eff}=\frac{\hbar}{2\pi}\int_{-\infty}^{\infty} dx\left[v_{J}(\partial_{x}\hat{  \theta})^2+\tilde{v}_{N}e^{-i\delta_{\gamma}}(\partial_{x}\hat{ \phi})^2\right].
}

To analyze the Hamiltonian (\ref{effective}), we first  perform the mode expansions and reduce the problem to that of complex harmonic potentials \cite{RSK02,AJ03}. We then analytically obtain the exact effective ground state $|\Psi_{g,\gamma}\rangle$ and show that it has the lowest real part of the  eigenenergy and the longest lifetime among all the eigenstates. This indicates that the effective ground state survives longest in the post-selected dynamics.  In the limit of $\gamma\to 0$, $|\Psi_{g,\gamma}\rangle$ reduces to the ordinary ground state of the TLL model \cite{FS11}. Finally, we calculate the correlation functions of $|\Psi_{g,\gamma}\rangle$ and show that a unique critical behavior emerges as a consequence of the measurement backaction. 

\subsection{Ground-state wave function}

We here obtain the ground state and the spectrum of the effective non-Hermitian TLL Hamiltonian (\ref{effective}). We assume $0\leq \delta_{\gamma}<\pi/2$ so that there exists a metastable ground state (see the discussion below). We perform the mode expansions of the fields $\hat{ \phi}(x)$ and $\hat{  \theta}(x)$:
\eqn{
\hat{ \phi}(x)&\!=\!&\!\!-\!\!\sum_{k\neq 0}\!i\!\!\cdot\!{\rm sgn}(k)\!\sqrt{\frac{\pi\tilde{K}_{\gamma}}{2L|k|}}\!e^{-\alpha|k|/2-ikx}(\hat{b}_{k}^{\dagger}\!\!+\!\hat{b}_{-k}),\\
\hat{  \theta}(x)&=&\sum_{k\neq 0}i\sqrt{\frac{\pi}{2\tilde{K}_{\gamma}L|k|}}e^{-\alpha|k|/2-ikx}(\hat{b}_{k}^{\dagger}-\hat{b}_{-k}),
}
where $L$ is the system size, $\tilde{K}_{\gamma}$ is the renormalized TLL parameter $\tilde{K}_{\gamma}\equiv\sqrt{v_{J}/\tilde{v}_{N}}$, $\hat{b}_{k}$ ($\hat{b}^{\dagger}_{k}$) annihilates (creates) a mode with  momentum $k=2\pi m/L$ ($m=\pm1,\pm2,\ldots$), $\alpha\to +0$ is a short-distance cutoff, and we ignore zero modes which are irrelevant to the following  discussion about the ground state. The effective Hamiltonian (\ref{effective}) is then rewritten as
\eqn{
\hat{H}_{\rm eff}&=&\frac{\hbar\tilde{v}}{4}\sum_{k\neq 0}|k|
\begin{pmatrix}
\hat{b}_{k}^{\dagger}&\hat{b}_{-k}
\end{pmatrix}
\begin{pmatrix}
e^{-i\delta_{\gamma}}+1&e^{-i\delta_{\gamma}}-1\\
e^{-i\delta_{\gamma}}-1&e^{-i\delta_{\gamma}}+1
\end{pmatrix}
\begin{pmatrix}
\hat{b}_{k}\\ \hat{b}_{-k}^{\dagger}
\end{pmatrix}\nonumber\\  \label{effh2}
&=&\hbar\tilde{v}\sum_{k>0}k\Biggl[\frac{e^{-i\delta_{\gamma}+1}}{2}(\hat{b}_{k}^{\dagger}\hat{b}_{k}+\hat{b}_{-k}^{\dagger}\hat{b}_{-k})\nonumber\\
&&\;\;\;\;\;\;\;\;\;\;+\frac{e^{-i\delta_{\gamma}}-1}{2}(\hat{b}_{k}\hat{b}_{-k}+\hat{b}_{k}^{\dagger}\hat{b}^{\dagger}_{-k})\Biggr],
}
where $\tilde{v}\equiv\sqrt{v_{J}\tilde{v}_{N}}$. In analogy with a quantized harmonic oscillator, we introduce the position and momentum operators $\hat{x}_{k}$ and $\hat{p}_{k}$ via
\eqn{
\hat{b}_{k}&=&\frac{\hat{x}_{k}+i\hat{p}_{k}}{\sqrt{2}},\\
\hat{b}_{k}^{\dagger}&=&\frac{\hat{x}_{k}-i\hat{p}_{k}}{\sqrt{2}},
}
and use them to rewrite Eq. (\ref{effh2}) as
\eqn{
\hat{H}_{\rm eff}&=&\hbar\tilde{v}\sum_{k>0}k\Biggl[\frac{e^{-i\delta_{\gamma}}+1}{4}(\hat{x}_{k}^2+\hat{p}_{k}^2+\hat{x}_{-k}^2+\hat{p}_{-k}^2)\nonumber\\
&&\;\;\;\;\;\;\;\;\;\;\;\;+\frac{e^{-i\delta_{\gamma}}-1}{2}(\hat{x}_{k}\hat{x}_{-k}-\hat{p}_{k}\hat{p}_{-k})\Biggr].\label{temp0}
}
We further introduce the center-of-mass and relative coordinates and momenta of the modes with $\pm k$, $\hat{x}_{k,\pm}$ and $\hat{p}_{k,\pm}$ via

\eqn{
\hat{x}_{k}&=&\frac{\hat{x}_{k,+}+\hat{x}_{k,-}}{\sqrt{2}},\;\; \hat{x}_{-k}=\frac{\hat{x}_{k,+}-\hat{x}_{k,-}}{\sqrt{2}},\label{temp1} \\
\hat{p}_{k}&=&\frac{\hat{p}_{k,+}+\hat{p}_{k,-}}{\sqrt{2}},\;\; \hat{p}_{-k}=\frac{\hat{p}_{k,+}-\hat{p}_{k,-}}{\sqrt{2}},\label{temp2}
}
where $k$ is a positive discrete momentum, i.e., $k=2\pi m/L$ with $m=1,2,\ldots$. Substituting Eqs. (\ref{temp1}) and (\ref{temp2}) into Eq. (\ref{temp0}), we arrive at the following Hamiltonian:
\eqn{
\hat{H}_{\rm eff}=\hbar\tilde{v}\sum_{k>0}\!k\!\left(\!\frac{e^{-i\delta_{\gamma}}}{2}\hat{x}^2_{k,+}\!\!+\!\!\frac{1}{2}\hat{p}_{k,+}^{2}+\frac{1}{2}\hat{x}_{k,-}^{2}\!\!+\!\!\frac{e^{-i\delta_{\gamma}}}{2}\hat{p}_{k,-}^{2}\!\!\right).\nonumber\\
}
The problem thus reduces to solving a set of non-Hermitian harmonic oscillators \cite{RSK02,AJ03}. We choose the basis in which the operators $\hat{x}_{k,+}$ and $\hat{p}_{k,-}$ are diagonalized so that the field $\hat{ \phi}(x)$ is also diagonalized as shown below. The metastable ground state $|\Psi_{g,\gamma}\rangle$ with the lowest eigenenergy can then be obtained as

\eqn{\label{ground}
\langle \{x_{k,+},p_{k,-}\}|\Psi_{g,\gamma}\rangle \!\propto \exp\left[-\frac{e^{-i\delta_{\gamma}/2}}{2}\sum_{k>0}(x_{k,+}^2+p_{k,-}^2)\right],\nonumber\\
}
which is a generalization of the ground-state wave function of a TLL \cite{FS11} to a non-hermitian case.
The eigenvalues can be obtained as 
\eqn{\label{preee1}\hbar\tilde{v}e^{-i\delta_{\gamma}/2}\sum_{k>0}k(n_{k,+}+n_{k,-}+1),} 
where $n_{k,\pm}$ are nonnegative integers labeling the eigenstates of the modes $(k,\pm)$. Since we assume $0\leq\delta_{\gamma}<\pi/2$, the energies are bounded from below and the ground-state wave function can be normalized. We note that, while the wave function (\ref{ground}) remains normalizable for $\pi/2\leq\delta_{\gamma}<\pi$, it can be shown that the full spectrum including the zero-mode contributions is no longer bounded from below in this regime. The negative imaginary part of eigenvalues found in Eq. (\ref{preee1}) indicates a finite lifetime of the eigenstate. We note that the ground state ($n_{k,\pm}=0$) has the lowest energy and the minimal imaginary part; thus, the ground state has the longest lifetime among all the eigenstates. Thus, the ground state discussed here has a clear dynamical meaning--- it survives  the longest in the post-selected dynamics.

We note that if $\delta_{\gamma}\geq\pi$, the energies are not bounded from below and there are no eigenstates having discrete eigenvalues because the wave function cannot be normalized (see Eq. (\ref{ground})).  The system is therefore dynamically unstable, and there is no metastable ground state for $\delta_{\gamma}\geq\pi$.

\subsection{Correlation function of the $\hat{ \phi}$ field}
 
 We first consider the correlation function $\langle e^{2i\hat{ \phi}(x)}e^{-2i\hat{ \phi}(y)}\rangle$, where $\langle\cdots\rangle$ denotes the expectation value with respect to $|\Psi_{g,\gamma}\rangle$. By performing the transformations described above, the mode expansion of the field $\hat{ \phi}(x)$ can be rewritten as
\eqn{
\hat{ \phi}(x)&=&i\sum_{k>0}\sqrt{\frac{\pi\tilde{K}_{\gamma}}{2Lk}}\Bigl[(\hat{x}_{k,+}+i\hat{p}_{k,-})e^{ikx}\nonumber\\
&&\;\;\;\;\;\;\;\;\;\;\;\;\;\;\;\;\;\;\;\;\;-(\hat{x}_{k,+}-i\hat{p}_{k,-})e^{-ikx}\Bigr].
}
Let $|\{ \phi_{k}\}\rangle$ be a simultaneous eigenstate of $\hat{ \phi}(x)$ $(0\leq x<L)$ satisfying
\eqn{
\hat{ \phi}(x)|\{ \phi_{k}\}\rangle&=& \phi(x)|\{ \phi_{k}\}\rangle,\\
 \phi(x)&=&\sqrt{\frac{\pi}{L}}\sum_{k>0}\left( \phi_{k}e^{ikx}+ \phi_{k}^{*}e^{-ikx}\right),
}
where $ \phi_{k}$ is related to the eigenvalues of the operators $\hat{x}_{k,+}$ and $\hat{p}_{k,-}$ via
\eqn{\label{pre}
i(x_{k,+}+ip_{k,-})=\sqrt{\frac{2k}{\tilde{K}_{\gamma}}} \phi_{k}.
}
The ground state in Eq. (\ref{ground}) can then be expressed as
\eqn{
\langle \{ \phi_{k}\}|\Psi_{g,\gamma}\rangle=\frac{1}{\sqrt{\mathcal{N}}}\exp\left(-\frac{e^{-i\delta_{\gamma}/2}}{\tilde{K}_{\gamma}}\sum_{k>0}k| \phi_{k}|^2\right),
}
where $\mathcal{N}$ is a normalization constant. The correlation function can be expressed as
\eqn{
&&\langle e^{2i\hat{ \phi}(x)}e^{-2i\hat{ \phi}(y)}\rangle=\!\frac{1}{\mathcal{N}}\!\int\!\!\mathcal{D} \phi\mathcal{D} \phi^{*}\times\nonumber\\
&&\exp\Biggl\{\sum_{k>0} \Biggl[\!-\!\frac{2k\cos(\delta_{\gamma}/2)}{\tilde{K}_{\gamma}}| \phi_{k}|^2
+2i\sqrt{\frac{\pi}{L}} \phi_{k}(e^{ikx}-e^{iky})\nonumber\\
&&\;\;\;\;\;\;\;\;\;\;\;\;\;\;\;\;\;\;\;\;\;\;\;\;+2i\sqrt{\frac{\pi}{L}} \phi_{k}^{*}(e^{-ikx}-e^{-iky})\Biggr]\Biggr\}.
}
After performing the Gaussian integrations, we obtain
\eqn{
\langle e^{2i\hat{ \phi}(x)}e^{-2i\hat{ \phi}(y)}\rangle\!\!=\!\!\exp\!\left[\!-\frac{\tilde{K}_{\gamma}}{\cos(\delta_{\gamma}/2)}\!\sum_{k>0}\!\!\frac{k_{1}}{k}\!(2\!\!-\!\!e^{ikr}\!\!-\!\!e^{-ikr})\right],\nonumber\\
}
where we define $k_{1}\equiv 2\pi/L$ and $r\equiv x-y$. Here, the sum over $k>0$ can be taken with a regularization trick:
\eqn{
\sum_{k>0}\!\!\frac{k_{1}e^{-\alpha k}}{k}(2\!-\!e^{ikr}\!\!-\!e^{-ikr})\!\to\! -2\!\ln\left(\!\frac{\alpha k_{1}}{2\sin(k_{1}r/2)}\!\!\right)\!,
}
where we take the limit of $\alpha k_{1}\ll 1$.
We thus obtain the critical behavior of the correlation function as
\eqn{
\langle e^{2i\hat{ \phi}(x)}e^{-2i\hat{ \phi}(y)}\rangle&=&\left(\frac{\alpha}{(L/\pi)\sin\left(\pi r/L\right)}\right)^{\frac{2\tilde{K}_{\gamma}}{\cos(\delta_{\gamma}/2)}}\nonumber\\
&\to& \left(\frac{\alpha}{r}\right)^{2K_{ \phi}},
}
where we take the limit of $L\gg r$ and introduce the critical exponent $K_{ \phi}$ by
\eqn{\label{kphi}
K_{ \phi}(\gamma)=\frac{\tilde{K}_{\gamma}}{\cos\left(\frac{\delta_{\gamma}}{2}\right)}.
}

\subsection{Correlation function of the  $\hat{  \theta}$ field}

Let us next consider the correlation function $\langle e^{i\hat{  \theta}(x)}e^{-i\hat{  \theta}(y)}\rangle$. Since the operator $\hat{\theta}$ can be expanded as
\eqn{
\hat{  \theta}(x)=-i\sum_{k>0}\sqrt{\frac{\pi}{2\tilde{K}_{\gamma}Lk}}\Bigl[i\hat{p}_{k,+}(e^{ikx}+e^{-ikx})\nonumber\\
+\hat{x}_{k,-}(e^{ikx}-e^{-ikx})\Bigr],
} 
it acts on an eigenstate of the operators $\hat{x}_{k,+}$ and $\hat{p}_{k,-}$ as 
\eqn{
&&e^{i\hat{  \theta}(x)}|\{x_{k,+},p_{k,-}\}\rangle=\nonumber\\
&&\;\;\;\;\;\;\;\;\Biggl|\Biggl\{x_{k,+}-i\sqrt{\frac{\pi}{2\tilde{K}_{\gamma}Lk}}(e^{ikx}+e^{-ikx}),\nonumber\\
&&\;\;\;\;\;\;\;\;\;\;\;\;\;\;\;\;\;\;\;\;\;p_{k,-}+\sqrt{\frac{\pi}{2\tilde{K}_{\gamma}Lk}}(e^{ikx}-e^{-ikx})\Biggr\}\Biggr\rangle.
}
Using Eq. (\ref{pre}), we can rewrite this as
\eqn{
e^{i\hat{  \theta}(x)}|\{ \phi_{k}\}\rangle=\biggl|\biggl\{ \phi_{k}-\frac{i}{k}\sqrt{\frac{\pi}{L}}e^{-ikx}\biggr\}\biggr\rangle.
}
Using this result, we can express the correlation function as
\eqn{
&&\langle e^{i\hat{  \theta(x)}}e^{-i\hat{  \theta}(y)}\rangle=\frac{1}{\mathcal{N}}\int\mathcal{D} \phi\mathcal{D} \phi^{*}\times\nonumber\\
&&\exp\Biggl[-\frac{1}{\tilde{K}_{\gamma}}\sum_{k>0}k\Biggl(e^{i\delta_{\gamma}}\Bigl| \phi_{k}-\frac{i}{k}\sqrt{\frac{\pi}{L}}e^{-ikx}\Bigr|^2\nonumber\\
&&\;\;\;\;\;\;\;\;\;\;\;\;\;\;\;\;\;\;\;\;\;\;\;\;\;\;\;\;+e^{-i\delta_{\gamma}}\Bigl| \phi_{k}-\frac{i}{k}\sqrt{\frac{\pi}{L}}e^{-iky}\Bigr|^2\Biggr)\Biggr].
}
After performing the Gaussian integrations, we obtain 
\eqn{
&&\langle e^{i\hat{  \theta(x)}}e^{-i\hat{  \theta}(y)}\rangle
\nonumber\\
&=&\exp\left[-\frac{\pi}{2\tilde{K}_{\gamma}L\cos(\delta_{\gamma}/2)}\sum_{k>0}\frac{1}{k}(2\!-\!e^{ikr}\!-\!e^{-ikr})\!\right]\\
&=&\left(\!\frac{\alpha}{(L/\pi)\sin\left(\pi r/L\right)}\!\right)^{\frac{1}{2\tilde{K}_{\gamma}\!\cos(\delta_{\gamma}/2)}}\!\to\! \left(\frac{\alpha}{r}\right)^{1\!/(2K_{  \theta}\!)},
}
where we take the limit of $L\gg r$ and define the critical exponent $K_{  \theta}$ by
\eqn{\label{ktheta}
K_{  \theta}(\gamma)=\tilde{K}_{\gamma}\cos\left(\frac{\delta_{\gamma}}{2}\right).
}

\subsection{Shifts in critical exponents due to measurement backaction}

Let us discuss the critical properties of the effective ground state $|\Psi_{g,\gamma}\rangle$. Physically,  the effective TLL parameters $K_{\theta, \phi}$ introduced by Eqs. (\ref{ktheta}) and (\ref{kphi})  describe the critical behavior of the one-particle correlation and the density correlation, respectively:
\eqn{ 
\langle\hat{\Psi}^{\dagger}(r)\hat{\Psi}(0)\rangle&\propto&\left(\frac{1}{r}\right)^{1/(2K_{  \theta})}, \\
\langle\hat{\rho}(r)\hat{\rho}(0)\rangle \!- \!\rho_{0}^2 \!&=& -\frac{K_{ \phi}}{2\pi^2 r^2}\!+\!{\rm const.}\times\frac{\cos(2\pi\rho_{0}r)}{r^{2K_{ \phi}}}.
}
The two characteristic parameters $K_{  \theta}$ and $K_{\phi}$ mark a unique critical behavior beyond the realm of the standard TLL universality class--- in the latter, the single TLL  parameter $K$ governs the critical properties. We note that $K_{  \theta}$ can be measured by interfering Bose gases \cite{PA06,SH08}, while  $K_{ \phi}$ can be found by analyzing $\it in$-$\it situ$ density fluctuations  \cite{SHF10}.

\begin{figure}[t]
\includegraphics[width=86mm]{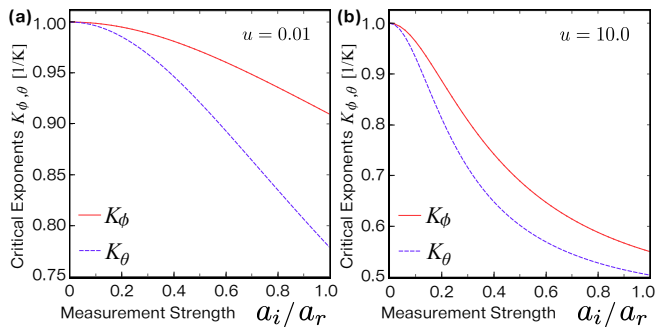}
\caption{\label{fig4}
(color online). Measurement-induced shifts of the critical exponents. The effective TLL parameters $K_{ \phi}$ (red solid curve) and $K_{  \theta}$ (blue dashed curve) are plotted against the ratio of the inelastic scattering length $a_{i}$ to the elastic one $a_{r}$ for the normalized strength of the interaction (see below Eq. (\ref{asp})) $u=0.01$ and $10.0$ in (a) and (b), respectively.
}
\end{figure}
 
Figure \ref{fig4} shows the shifts of the critical exponents $K_{ \phi,  \theta}$ as functions of the normalized measurement strength $\gamma/g=a_i/a_r$. Here we consider both cases of weak and strong interactions and perform the analytic continuation of the corresponding asymptotic expressions (\ref{asp}) through the replacement $g\to g-i\gamma$ to relate the measurement strength $\gamma$ to the renormalized parameters $\tilde{v}_{N}$ and $\delta_{\gamma}$. The decrease in $K_{ \phi,  \theta}$ can be interpreted as an effective enhancement of the interaction strength due to the  contribution to its imaginary part from the measurement backaction. Physically, such an enhanced interaction arises from the increased repulsion between atoms due to the continuous quantum Zeno effect \cite{BA00,FP02}. The split between the two characteristic parameters $K_{ \phi,  \theta}$ arises from the additional degree of freedom $\delta_{\gamma}$ in the parameter space of the effective Hamiltonian in Eq. (\ref{effective}) and is a manifestation of the non-Hermiticity of the underlying model. This unique feature indicates a dramatic departure from the conventional TLL behavior due to the measurement backaction. These findings should also be relevant to various 1D critical systems in addition to 1D Bose gas owing to the universality of the effective Hamiltonian (\ref{effective}). A change in the critical exponent in the non-Hermitian system can also be found in the steady-state transition \cite{LTE14}, while its universality as  critical phenomena remains to be understood.  

 \section{Experimental implementations\label{expsec}}
 
\begin{figure*}[t]
\includegraphics[width=160mm]{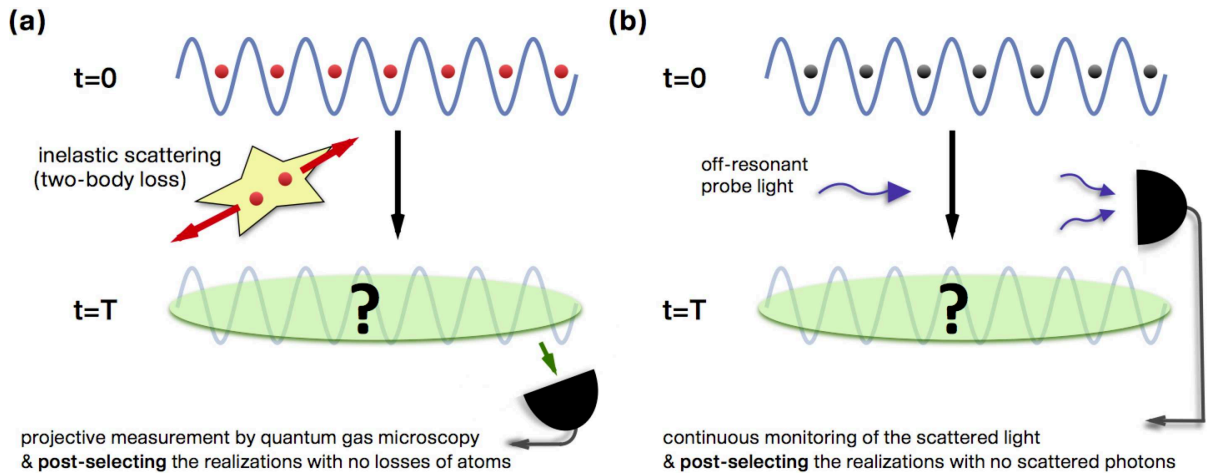}
\caption{\label{fig5}
Schematic figures illustrating the methods to implement the post-selection processes and simulate the non-Hermitian dynamics in which the system undergoes (a) inelastic two-body loss and (b) continuous monitoring of an off-resonantly scattered photons. In (a), we first prepare an initial state with a well-estimated total particle number, and let the system evolve in time. If an inelastic scattering process occurs during the time evolution, this leads to a loss of pairs of atoms. At the final stage, we perform the projective measurement by using quantum gas microscopy and post-select the realizations in which no losses of atoms occur. In (b), after preparing an initial state, we let the system evolve in time and simultaneously probe the system by irradiating an off-resonant light on the system and continuously monitor the scattered light. Then, after a duration of time $T$, we post-select the realizations in which  no scattered photons are observed.
}
\end{figure*}

Here we discuss how we can implement post-selection processes to simulate the non-Hermitian dynamics in ultracold atoms. We propose the following two schemes: (A) use of a system subject to an inelastic two-body loss of atoms combined with measurement of the atom number with quantum gas microscopy to post-select the realizations in which no quantum jumps (atomic losses) occur, and (B) irradiation of an off-resonant light combined with continuous monitoring of the scattered photons. We also estimate experimental parameters for each proposal.

\subsection{Post-selection by the atom-number measurement with quantum gas microscopy}
We consider atomic systems accompanying an inelastic scattering process leading to a two-body loss of atoms. We first prepare a desired initial state of atoms trapped in an optical lattice. The current techniques can now prepare almost an arbitrary configuration of atoms with an accurate estimated total number of particles at the single-particle level \cite{IR15}. A possible loss of atoms during the preparation can be circumvented by loading atoms in a stable state and then transferring them  into a metastable state subject to an inelastic scattering process. We then let the system evolve in time; if an inelastic scattering between two atoms occurs during the time evolution, then a pair of atoms is lost  from an optical potential because colliding atoms acquires a much larger amount of kinetic energies than the potential energy (Fig. \ref{fig5}(a)). Thus, such a process can effectively be described by the operator $\hat{M}_{i}=\hat{b}_{i}^{2}$. 

Then, after some duration time, we perform a projection measurement of site-occupation  numbers by using quantum gas microscopy \cite{PPM152}. We compare the total number of particles in the final state with that in the initial state, and post-select the realizations in which the two numbers agree (Fig. \ref{fig5}(a)). (If they disagree, quantum jumps due to two-body loss should have occurred and the time evolution of the system cannot be described by the non-Hermitian Hamiltonian (\ref{nonher}).) In this way, we can simulate the dynamics governed by the effective non-Hermitian Hamiltonian. Here the system can be interpreted as being subject to the continuous measurement backaction from an environment that causes the two-body losses of atoms. The measurement strength $\gamma$ can be determined by conducting experiments {\it without} performing post-selections and measuring the loss rate of atoms. 

\subsection{Post-selection by continuous monitoring of off-resonantly scattered photons}
In this scheme, we use an off-resonant probe light and continuously monitor the scattered photons (Fig. \ref{fig5}(b)).   Specifically, we proceed as follows. First, we prepare an initial state, and  then shine an off-resonant probe light on the system. In this measurement process,  light-induced inelastic collisions are significantly suppressed owing to far detuning, and the dominant process becomes a photon scattering process \cite{YA15}. In such a situation, it is known that the measurement process can be described by the operator $\hat{M}_{i}=\hat{n}_{i}$ (see e.g., Ref. \cite{PH10}). Physically speaking, this operator form reflects the fact that the atom occupation number can be determined by collecting scattered photons and analyzing their interference patterns. We then continuously monitor the scattered light by, for example, collecting photons by a high numerical aperture lens and detecting them with a high-sensitivity CCD imaging device, as implemented by quantum gas microscopy.  During the continuous monitoring, we  post-select the realizations in which no photons are detected. In this way, ideally, we can simulate the dynamics governed by the non-Hermitian Hamiltonian in Eq. (\ref{nonher}) with the operator $\hat{M}_{i}=\hat{n}_{i}$. In practice,  however, we can collect only a portion of the scattered photons and our results can be tested if a possible heating caused by the undetected photons does not smear our theoretical predictions for appropriately chosen experimental parameters as discussed below. 

\subsection{Experimental parameters}
Here we discuss the feasibilities of the two proposals discussed above by estimating some experimental parameters. Let us first consider the experimental feasibility of the proposal (A). A system accompanying an inelastic two-body loss can be realized by using, e.g., the metastable $^3P_{2}$ state of $^{174}$Yb atoms having an inelastic scattering length $a_{i}=2.8\;{\rm nm}$ in addition to the elastic one $a_{r}=5.8\;{\rm nm}$ \cite{US12}. Also, one may use polar molecules, which have inelastic scattering channels resulting in two-body losses, or the light-induced inelastic collisions between stable atoms. The dominant factor in determining the experimental fidelity of post-selections is the detection fidelity of atom-number measurement with quantum gas microscopy. In fact, it is known that the fidelity of quantum gas microscopy is very high and almost reaches the near-unit fidelity (99.5\% in Ref. \cite{SJF10}). Indeed, such a high fidelity has  already been sufficient to enable experimenters to implement the post-selections to, for example, reduce the entropy of the system and eliminate a possible experimental error \cite{EM11,EM13,FT15,IR15}. While the technique has originally been restricted to the parity measurement, this restriction has been relaxed by recent experimental developments \cite{PPM152}. Since the fidelity of selecting null-outcome events decreases as the total number of atoms increases, the atom number should be prepared in a relatively small number such as several tens of atoms as demonstrated in Ref. \cite{BWS10}, where an experimental error can be made smaller than our predicted value of the relative shift of the transition point.  To further improve the experimental fidelity, a recent technique of the super-resolved observation \cite{AA15,YA15} can be useful.

We next consider the proposal (B). In this case,  the detection fidelity of  scattered photons is limited in general and the recoil energies of  undetected photons inevitably induce  heating of the system, which may smear out the predicted shifts of the critical point and the critical exponents. To discuss the appropriate experimental parameters, let us estimate the expected finite-temperature effect due to heating. Let $\eta$ be  the detection efficiency of the scattered photons and $\gamma$ be the scattering rate of photons by an atom and $\tau$ be the duration of the time evolution. Then, the net heating energy per atom caused by the recoil energies of the undetected scattered photons can be estimated to be $\delta E=(1-\eta)\gamma\tau\times\hbar^{2}k^{2}/(2m)$, where $k=2\pi/\lambda$ is the wavenumber of a scattered light and $m$ is the atomic mass. In general, it is known that a finite-temperature effect shifts the transition point \cite{ST10}, and thus our predicted shift is testable if the heating is not too large to mask the predicted shift. To be specific, we consider $^{87}$Rb atoms and an off-resonant light with wavelength $\lambda=1064\;{\rm nm}$. For the sake of concreteness, we set the measurement strength to be $\gamma/U=0.2$ at which the predicted shift of the transition point is about 10\% as shown in Fig. 2. As a typical duration time, we take the characteristic relaxation time scale $\tau\sim 1/\gamma$. According to Ref. \cite{ST10}, the temperature that induces the same amount of shift ($\sim10\%$) is $T_{\rm th}/J\simeq 3$ at $U/J\simeq 25$. Here we set $k_{\rm B}=1$ and $J/\hbar=3$ms as the hopping rate. Combining these experimental parameters, the condition $\delta E<T_{\rm th}/J$ that the predicted shift of the transition point is not masked by the finite-temperature effect imposes a constraint on the detection fidelity $\eta>0.08$. Such a condition can be met by, for example, a high detection efficiency of photons which can be achieved with quantum gas microscopy ($\eta=0.1\sim0.2$ in Refs. \cite{BWS09,SJF10,CLW15,PMF15,EH15,OA15,EGJ15,PPM152}). We however note that a larger detection efficiency is required in order for the heating effect to be an order of magnitude smaller than the predicted shift of the transition point. 
We next estimate the experimental benchmark to test our result on the variations of the critical exponents. In this case, the heating caused by the undetected photons is characterized by a finite thermal correlation length $\xi_{T}\equiv \hbar^{2}\rho_{0}\pi/(mT)$, where $\rho_{0}$ is the number density of atoms. If the heating is not too large and a length scale of interest $r$ in the correlation functions satisfies $r<\xi_{T}$, then the critical behavior can be observed and the associated exponents should  experimentally be determined by measuring the correlation functions. In addition, we also need to consider the condition to ensure the validity of the Tomonaga-Luttinger-liquid (TLL) low-energy description. According to Ref. \cite{KKV05},  such conditions in weakly and strongly interacting regimes are
\eqn{
\frac{2mT}{\hbar^{2}\rho_{0}^{2}u^{2}}&\lesssim& 10^2\;\;\;{\rm (weakly\; interacting\; regime, \;{\it u}=0.01)},\nonumber\\ \\
\frac{2mT}{\hbar^{2}\rho_{0}^{2}u^{2}}&\lesssim& 10^{-2}\;\;\;{\rm (strongly \;interacting\; regime, \;{\it u}=10)},\nonumber\\ \label{ineqe}
}
where $u$ is the dimensionless interaction parameter defined in Sec. \ref{criticalsec}. To be specific, let us consider the above experimental situations with the density $\rho_{0}=55\;\mu$m$^{-1}$ and a length scale $r=10\;\mu$m. Then, in the weakly interacting regime, the second condition on the validity of the TLL description is more crucial and leads to the constraint on the detection efficiency $\eta>0.13$. This condition is within the reach of the current experimental techniques of quantum gas microscopy as mentioned above. In contrast, in the strongly interacting regime, owing to the large value of $u$, the low-energy description is more robust against the finite-temperature effect, as inferred from Eq. (\ref{ineqe}), and the experimental constraint is much less stringent than the weakly interacting case.

\section{Conclusions\label{consec}}
We have investigated how the notions of quantum phase transitions and universality in quantum critical phenomena can be extended to many-body systems subject to the measurement backaction of continuous observation. We have introduced effective non-Hermitian Hamiltonians of interacting bosons in an optical lattice and a one-dimensional trap, and analyzed their effective ground states, i.e., the states having the lowest real parts of eigenvalues. It is shown that the effective ground states in both models have the minimal imaginary parts of eigenvalues and thus survive longest in the non-Hermitian dynamics. In the former model, by performing the mean-field and strong-coupling-expansion analyses, we have shown that the measurement backaction can shift the superfluid--to-Mott-insulator transition point and expand the Mott lobes. In the latter model, we have derived the low-energy effective field theory that describes critical behavior of the one-dimensional Bose gas subject to the measurement backaction, and  found the new critical exponents that depend on the strength of the measurement. This indicates a unique critical behavior beyond the realm of the standard Tomonaga-Luttinger liquid universality class. Owing to the universality of the effective field theory, our findings should also be relevant to other one-dimensional critical systems.
To test our predictions, we have discussed two experimental schemes in ultracold atoms by using quantum gas microscope, and also estimated possible experimental parameters. In view of recent developments in engineering non-Hermitian Hamiltonians \cite{GT15,ZB15}, it seems of interest to explore unique aspects of non-Hermitian systems such as topological structures around exceptional points \cite{IR09} and spectral singularity in $\mathcal PT$ symmetric systems \cite{BCM98,EMG08} in the context of many-body physics, especially in critical regimes.

\begin{acknowledgements}
We acknowledge fruitful discussions with T. Fukuhara, Z. Gong, T. Tomita, Y. Takahashi, I. Danshita, and T. Mori. This work was supported by KAKENHI Grant Nos. JP25800225 and JP26287088 from the Japan Society for the Promotion of Science (JSPS), and a Grant-in-Aid for Scientific Research on Innovative Areas ``Topological Materials Science" (KAKENHI Grant No. JP15H05855), and the Photon Frontier Network Program from MEXT of Japan, ImPACT Program of Council for Science, Technology and Innovation (Cabinet Office, Government of Japan), and the Mitsubishi Foundation. Y. A. acknowledges support from the Japan Society for the Promotion of Science through Program for Leading Graduate Schools (ALPS) and Grant No. JP16J03613.
\end{acknowledgements}

\bibliography{reference}

\end{document}